# Double Stage Delay Multiply and Sum Beamforming Algorithm: Application to Linear-Array Photoacoustic Imaging

Moein Mozaffarzadeh, Ali Mahloojifar*, Mahdi Orooji, Saba Adabi and Mohammadreza Nasiriavanaki

*Abstract*—Photoacoustic imaging (PAI) is an emerging medical imaging modality capable of providing high spatial resolution of Ultrasound (US) imaging and high contrast of optical imaging. Delay-and-Sum (DAS) is the most common beamforming algorithm in PAI. However, using DAS beamformer leads to low resolution images and considerable contribution of off-axis signals. A new paradigm namely Delay-Multiply-and-Sum (DMAS), which was originally used as a reconstruction algorithm in confocal microwave imaging, was introduced to overcome the challenges in DAS. DMAS was used in PAI systems and it was shown that this algorithm results in resolution improvement and sidelobe degrading. However, DMAS is still sensitive to high levels of noise, and resolution improvement is not satisfying. Here, we propose a novel algorithm based on DAS algebra inside DMAS formula expansion, Double Stage DMAS (DS-DMAS), which improves the image resolution and levels of sidelobe, and is much less sensitive to high level of noise compared to DMAS. The performance of DS-DMAS algorithm is evaluated numerically and experimentally. The resulted images are evaluated qualitatively and quantitatively using established quality metrics including signal-to-noise ratio ($SNR$), full-width-half-maximum ($FWHM$) and contrast ratio ($CR$). It is shown that DS-DMAS outperforms DAS and DMAS at the expense of higher computational load. DS-DMAS reduces the lateral valley for about 15 $dB$ and improves the $SNR$ and $FWHM$ better than 13% and 30%, respectively. Moreover, the levels of sidelobe are reduced for about 10 $dB$ in comparison with those in DMAS.

## I. INTRODUCTION

PHOTOACOUSTIC imaging (PAI) is an emerging medical imaging modality developed over the past few years and provides structural, functional and anatomical information [1], [2]. This modality uses short laser pulse to generate Ultrasound (US) waves based on thermoelastic effect and detected waves are used to reconstruct the optical absorption distribution map of the tissue [3]. The main motivation for using PAI is combining the high spatial resolution of US imaging and high contrast of optical imaging in one single imaging modality [4]. PAI is a multiscale imaging modality used in different preclinical and clinical applications e.g., tumor detection [5],

*Corresponding author
M.Mozaffarzadeh, A.Mahloojifar and M.Orooji are with the Department of Biomedical Engineering, Tarbiat Modares University, Tehran, Iran (e-mail: moein.mfh@modares.ac.ir; mahlooji@modares.ac.ir; morooji@modares.ac.ir).
S.Adabi and M.Nasiriavanaki are with the Department of Biomedical Engineering, Wayne State University, 818 W. Hancock, Detroit, Michigan, USA (e-mail: saba.adabi@wayne.edu; mrn.avanaki@wayne.edu).

[6], ocular imaging [7], monitoring oxygenation in blood vessels [8], and functional imaging [9], [10]. There are two types of PAI: photoacoustic tomography (PAT) and photoacoustic microscopy (PAM) [11], [12]. In PAT, the induced acoustic waves are detected using an array of US transducers in the form of linear, arc or circular shape. Moreover, mathematical reconstruction algorithms are used to obtain optical absorption distribution map of the tissue [13]–[15]. There are inherent artifacts in photoacoustic (PA) reconstructed images caused by imperfect reconstruction algorithms. Reducing these artifacts has become a crucial challenge in PA image reconstruction for different number of transducers and different properties of imaging media [16]–[18].

Since PA images are obtained based on induced US signals, there are many similarities between PA and US image formation. Consequently, several US beamforming algorithms can be used in PA beamforming with some modifications [19]. These modifications have led to usage of different hardware to implement an integrated US-PA imaging device. There are many studies focused on using one beamforming technique for US and PA image formation to reduce the cost of system [20], [21]. Delay-and-Sum (DAS) can be considered as one of the most common beamforming algorithm in US imaging. However, it is a blind and non-adaptive beamformer which results in low resolution images with high levels of sidelobe [22]. The problem of a proper beamforming algorithm has been widely investigated in US imaging in different cases of study [23], [24]. The blindness of DAS algorithm can be addressed by extension of the receive aperture length in Phased Synthetic Aperture (PSA) imaging [25], and adaptive beamforming such as Minimum Variance (MV) [26]. MV adaptive beamformer, having considerable applications in RADAR and US imaging, has been modified for US imaging over the past few years in different fields of study such as complexity reduction [27], [28], US pixel-based beamforming [29], and 3D US imaging [30]. Apart from that, beamforming of plane-wave in the Fourier domain can be used to achieve fast and accurate image reconstruction [31]. Recently, Delay and Standard Deviation (DASD) beamforming algorithm was introduced in order to address the relatively poor appearance of interventional devices such as needles, guide wires, and catheters, in conventional US images [32]. To improve the quality of reconstructed images using DAS, explained in [33], Matrone et al. proposed a new beamforming algorithm namely Delay-Multiply-and-Sum (DMAS) which was used as a reconstruction algorithm in confocal microwave imaging for breast cancer detection [34]. DMAS has been used with Multi-Line Transmission (MLT) for high frame-rate US imaging [35]. Both DAS and DMAS beamformers calculate delays and samples for corresponding elements of array. However, in DMAS before summation, calculated samples are combinatorially coupled and multiplied. This algorithm was recently used in PAI and it was proved that it can effectively enhance the PA images in terms of sidelobe levels and resolution [36]. In this paper, a modified version of DMAS beamformer, namely Double Stage DMAS (DS-DMAS), is introduced. By expanding the DMAS algorithm, it is proved that DMAS can be manipulated by the summation of DAS terms and it is proposed to use DMAS algorithm instead of existing DAS algorithm for all the mathematical terms.

The rest of the paper is organized as follows. Section II contains the photoacoustic wave equations and the theory of beamformers. Proposed method is introduced in section III. Numerical simulation of imaging system and experimental design along with results and performance evaluation are presented in section IV and section V, respectively. Discussion is presented in section VI and finally the conclusion is presented in section VII.

## II. BACKGROUND

When PA signals are detected by linear array of US transducers, US beamforming algorithms such as DAS can be used to reconstruct the image from detected PA signals using following equation:

$$y_{DAS}(k) = \sum_{i=1}^{M} x_i(k - \Delta_i), \quad (1)$$

where $y_{DAS}(k)$ is the output of beamformer, $k$ is time index, $M$ is the number of array elements and $x_i(k)$ and $\Delta_i$ are detected signals and corresponding time delay for detector $i$, respectively [27]. DAS beamformer is the most common beamforming algorithm in US and PA imaging due to simple implementation and realtime imaging capability. However, this algorithm suffers from low level of off-axis signals rejection and weak suppression of interfering signals. Consequently, DAS results in reconstructed images having high levels of sidelobe and low resolution. To address incapabilities of DAS, DMAS was suggested in [33], [36]. The same as DAS, DMAS calculates corresponding samples for each element of array based on delays, but samples are multiplied before adding them up. The DMAS formula is given by:

$$y_{DMAS}(k) = \sum_{i=1}^{M-1} \sum_{j=i+1}^{M} x_i(k-\Delta_i) x_j(k-\Delta_j), \quad (2)$$

To overcome the dimensionally squared problem of (2), following modifications are suggested in [33]:

$$\hat{x}_{ij}(k) = \\ \text{sign}[x_i(k-\Delta_i) x_j(k-\Delta_j)] \sqrt{|x_i(k-\Delta_i) x_j(k-\Delta_j)|}, \quad (3) \\ \text{for} \quad 1 \leq i \leq j \leq M.$$

$$y_{DMAS}(k) = \sum_{i=1}^{M-1} \sum_{j=i+1}^{M} \hat{x}_{ij}(k). \quad (4)$$

Performing sign, absolute and square root after coupling procedure in (3) and (4), which requires $(M^2 - M)/2$ computations for each pixel, result in slow imaging. To put it more simply, sometimes these library functions require many clock cycles which leads to improper timing performance of DMAS algorithm. Applying following procedure to the received PA signals reduces computational number of the sign, absolute and square root operations to $M$ for each pixel [36]:

$$\bar{x}_i(k) = \text{sign}[x_i(k)]\sqrt{x_i(k)}, \quad \text{for} \quad 1 \leq i \leq M. \quad (5)$$

$$\hat{x}_{ij}(k) = \bar{x}_i(k)\bar{x}_j(k), \quad \text{for} \quad 1 \leq i \leq j \leq M. \quad (6)$$

The procedure of DMAS algorithm can be considered as a correlation process which uses the autocorrelation of the aperture. In other words, the output of this beamformer is the spatial coherence of detected PA signals and it is a non-linear beamforming algorithm.

## III. PROPOSED METHOD

In this paper, DMAS beamforming is proposed to be used instead of the existing DAS algorithm inside DMAS algebra. Initially, consider the expansion of DMAS algorithm, which can be written as follows:

$$y_{DMAS}(k) = \sum_{i=1}^{M-1} \sum_{j=i+1}^{M} x_{id}(k) x_{jd}(k) = \\ x_{1d}(k)\left[x_{2d}(k) + x_{3d}(k) + x_{4d}(k) + ... + x_{Md}(k)\right] \\ + x_{2d}(k)\left[x_{3d}(k) + x_{4d}(k) + ... + x_{Md}(k)\right] \\ + ... \\ + x_{(M-2)d}(k)\left[x_{(M-1)d}(k) + x_{Md}(k)\right] \\ + \left[x_{(M-1)d}(k) x_{Md}(k)\right], \quad (7)$$

where $x_{id}(k)$ and $x_{jd}(k)$ are delayed detected signals for element $i$ and $j$, respectively. According to (7), there is a DAS in each term of the expansion, which can be used to generate DS-DMAS beamformer as follows:

$$y_{DMAS}(k) = \sum_{i=1}^{M-1} \sum_{j=i+1}^{M} x_{id}(k) x_{jd}(k) = \\ \underbrace{\left[x_{1d}(k)x_{2d}(k) + x_{1d}(k)x_{3d}(k) + ... + x_{1d}(k)x_{Md}(k))\right]}_{\text{first term}} \\ + \underbrace{\left[x_{2d}(k)x_{3d}(k) + x_{2d}(k)x_{4d}(k) + ... + x_{2d}(k)x_{Md}(k)\right]}_{\text{second term}} \\ + ... \\ + \underbrace{\left[x_{(M-2)d}(k)x_{(M-1)d}(k) + x_{(M-2)d}(k)x_{Md}(k)\right]}_{\text{(M-2)th term}} \\ + \underbrace{\left[x_{(M-1)d}(k)x_{Md}(k)\right]}_{\text{(M-1)th term}}. \quad (8)$$

DMAS algorithm is a correlation process in which for each point of the image, calculated delays for each element of the array are combinatorially coupled and multiplied and the similarity of samples are obtained. In (8), there is a summation as a type of DAS algebra between all the terms. If the contribution of off-axis signals result in a high error in correlation process of DMAS, then summation of calculated correlations leads to summation of high range of error. It is proposed to use DMAS beamformer between each term of expansion instead of DAS, in which the samples go through another correlation procedure. To illustrate, consider the following equation:

$$y_{DS-DMAS}(k) = \sum_{i=1}^{M-2} \sum_{j=i+1}^{M-1} x_{it}(k) x_{jt}(k), \qquad (9)$$

where $x_{it}$ and $x_{jt}$ are the $i_{th}$ and $j_{th}$ term shown in (8). The expansion of DS-DMAS beamformer can be written as:

$$\begin{aligned} y_{DS-DMAS}(k) &= \sum_{i=1}^{M-2} \sum_{j=i+1}^{M-1} x_{it}(k) x_{jt}(k) = \\ & x_{1t}(k) \Big[ x_{2t}(k) + x_{3t}(k) + x_{4t}(k) + ... + x_{(M-1)t}(k) \Big] \\ &+ x_{2t}(k) \Big[ x_{3t}(k) + x_{4t}(k) + ... + x_{(M-1)t}(k) \Big] \\ &+ ... \\ &+ x_{(M-3)t}(k) \Big[ x_{(M-2)t}(k) + x_{(M-1)t}(k) \Big] \\ &+ \Big[ x_{(M-2)t}(k) x_{(M-1)t}(k) \Big]. \end{aligned} \qquad (10)$$

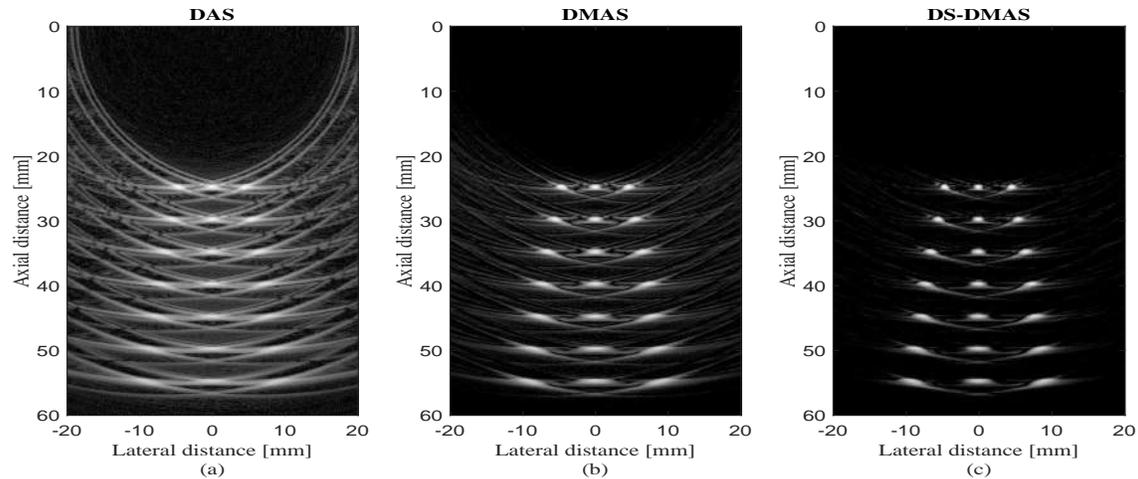

Fig. 1: Simulated three-point target using linear array. (a) DAS, (b) DMAS and (c) DS-DMAS. All images are shown with a dynamic range of 60 $dB$. 50 $dB$ noise was added to detected signals.

Since DAS is a non-adaptive beamformer and considers all calculated samples for each element of array identically, the acquired image by every term can blur the final reconstructed image. Using (10), blurring can be prevented and the noise of the reconstructed images can be reduced. Of note, the same procedure that was introduced in (5) and (6) in order to speed up the DMAS beamformer, is used in DS-DMAS for the same reason. In the section IV it is shown that DS-DMAS beamformer results in resolution improvement, and significant high level of noise and sidelobe reduction.

## IV. NUMERICAL RESULTS AND PERFORMANCE ASSESSMENT

In this section, numerical results are presented to illustrate the performance of the proposed algorithm in comparison with DMAS and DAS.

### A. Simulated Point Target

K-wave Matlab toolbox was used to simulate the numerical study [37]. Twenty-one 0.1 $mm$ spherical absorbers were positioned along the vertical axis every 5 $mm$ as initial pressure. The first absorber was 25 $mm$ away from the transducer surface. Each two absorbers at each depth were away 4.6 $mm$, 5.5 $mm$, 6.4 $mm$, 7.2 $mm$, 7.7 $mm$, 8.5 $mm$ and 9.1 $mm$ laterally from each other. Imaging region was 40 $mm$ in lateral axis and 60 $mm$ in vertical axis. Linear array having $M$=128 elements operating at 7 $MHz$ central frequency and 77 % fractional bandwidth was used to detect the PA signals generated from defined initial pressures. Speed of sound was assumed to be 1540 $m/s$ during simulations. 50 $dB$ and 10 $dB$ Gaussian noise was added to detected signals. Envelope detection, performed by means of the Hilbert transform, has been used in the end for all presented images, and the obtained lines are normalized and log-compressed to form the final images.

In Fig. 1, results of beamforming algorithms are shown, where 50 $dB$ noise is added to the detected signals. Fig. 1(a), Fig. 1(b) and Fig. 1(c) show the output of DAS, DMAS and DS-DMAS beamformer, respectively. The comparison of DMAS and DAS reveals that DMAS beamformer results in more distinguishable point targets in all depths of imaging, while DAS beamformer results in high levels of sidelobe and a low resolution image. In addition, point targets are hardly distinguishable after imaging depth of 35 $mm$ using DAS beamformer. By comparing reconstructed images using DS-DMAS and DMAS beamformers, it can be perceived that DS-DMAS leads to lower levels of sidelobe and more distinguished point targets. To compare the different beamforming algorithms in detail, the lateral variations of images formed by mentioned beamformers are presented in Fig. 2, where 50 $dB$ noise is added to the detected signals. Lateral variations in depth of 40 $mm$, 45 $mm$, 50 $mm$ and 55 $mm$ are shown in Fig. 2(a), Fig. 2(b), Fig. 2(c) and Fig. 2(d), respectively. It can be seen that sidelobe levels using DS-DMAS has the lowest level in comparison with two other algorithms in all depths of imaging, for instance, consider the depth of 55 $mm$ where sidelobe levels of DAS, DMAS and DS-DMAS are about -42 $dB$, -58 $dB$ and -72 $dB$, respectively. This conclusion can be perceived considering all the lateral variations in all depths of imaging. According to Fig .2, DS-DMAS beamformer leads to higher resolution in comparison with DAS and DMAS. In other words, by comparing the valley of lateral variations for mentioned beamformers, resolution elevation can be achieved. In particular, at depth of 50 $mm$, valley of lateral variation is reduced for about 30 $dB$, 51 $dB$ and 63 $dB$ for DAS, DMAS and DS-DMAS, respectively. Consequently, DS-DMAS results in more distinguished point targets and higher resolution in comparison to other beamformers. To conclude, in all depths of imaging DS-DMAS outperforms DAS and DMAS in cases of resolution and sidelobe levels. The three-point target simulation has been implemented in the presence of a high level of noise to evaluate the performance of beamformers in case of noise reduction. The reconstructed PA images are shown in Fig. 3, where 10 $dB$ Gaussian noise was added to the detected signals. Also, lateral variation of reconstructed images using noisy signals is shown in Fig. 4.

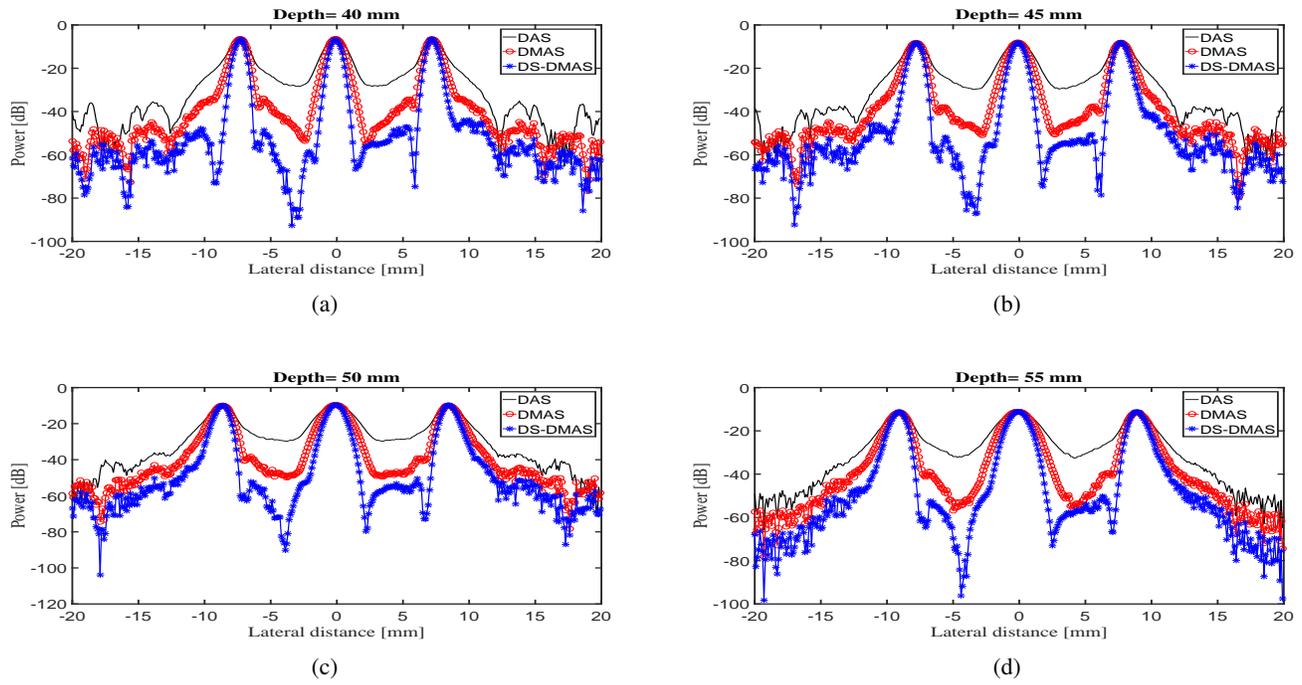

Fig. 2: Lateral variation of DAS, DMAS, DS-DMAS in depth of (a) 40 $mm$, (b) 45 $mm$, (c) 50 $mm$ and (d) 55 $mm$.

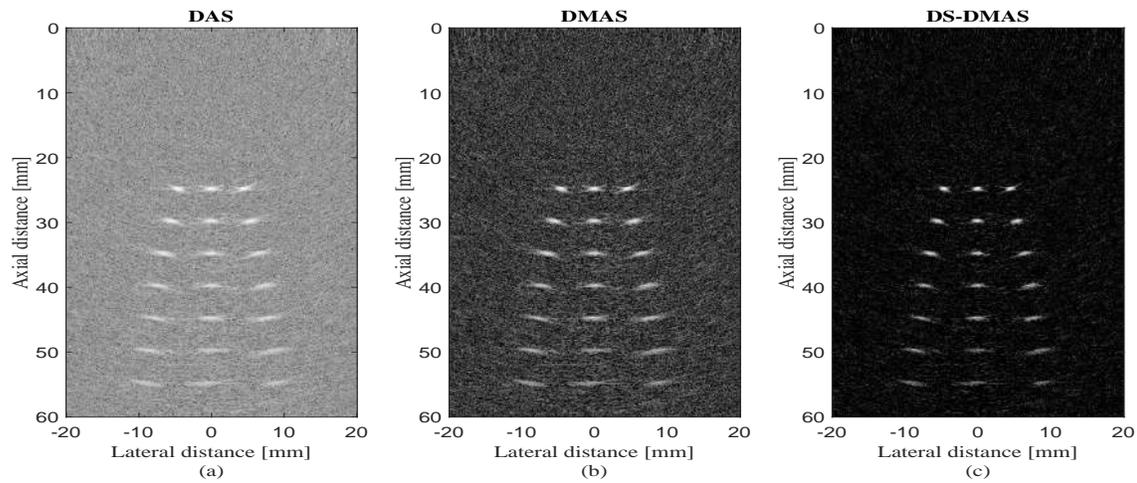

Fig. 3: Simulated three-point target using linear array. (a) DAS, (b) DMAS and (c) DS-DMAS. All images are shown with a dynamic range of 60 $dB$. 10 $dB$ noise was added to detected signals.

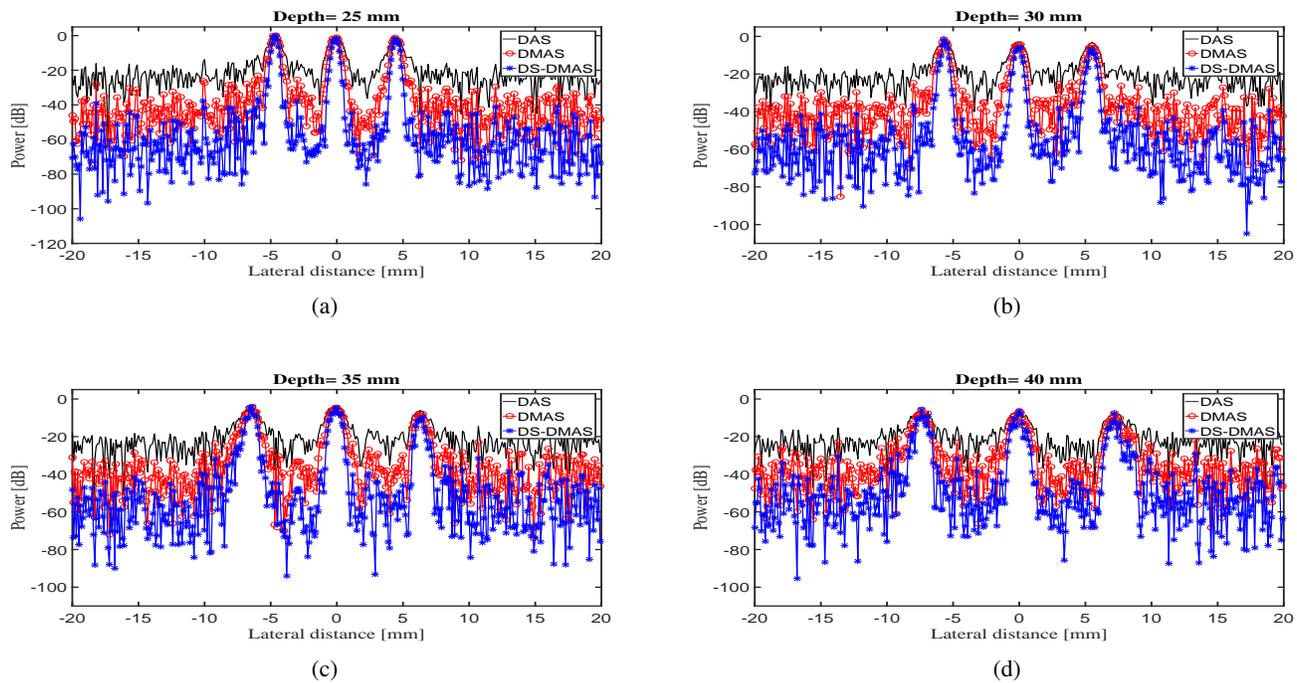

Fig. 4: Lateral variation of DAS, DMAS and DS-DMAS in depth of (a) 25 $mm$, (b) 30 $mm$, (c) 35 $mm$ and (d) 40 $mm$.

TABLE I: **SNR** ($dB$) Values in Different Depths.

| Depth($mm$) | 50dB noise | | |
|---|---|---|---|
| | DAS | DMAS | DS-DMAS |
| 25 | 34.8216 | 47.4809 | 53.6078 |
| 30 | 33.8332 | 44.9784 | 50.3456 |
| 35 | 33.0357 | 43.7758 | 48.8758 |
| 40 | 32.2885 | 42.0026 | 47.3446 |
| 45 | 31.1803 | 40.2071 | 45.335 |
| 50 | 30.1526 | 38.5532 | 43.7768 |
| 55 | 29.314 | 37.3118 | 42.9101 |

TABLE II: **FWHM** ($mm$) Values in Different Depths.

| Depth($mm$) | 50dB noise | | |
|---|---|---|---|
| | DAS | DMAS | DS-DMAS |
| 25 | 0.8613 | 0.5825 | 0.3931 |
| 30 | 1.0337 | 0.6299 | 0.463 |
| 35 | 1.165 | 0.8034 | 0.5288 |
| 40 | 1.3973 | 0.9735 | 0.6603 |
| 45 | 1.666 | 1.1767 | 0.8 |
| 50 | 1.985 | 1.4096 | 1.0356 |
| 55 | 2.2668 | 1.6922 | 1.2376 |

As can be seen in Fig. 3, the presence of noise is obvious in the reconstructed image using DAS beamformer and after depth of 45 $mm$ the point-targets are barely detectable. DMAS beamformer leads to more noise reduction and reconstructed image contains a lower levels of noise in comparison with DAS, but effect of noise still suffers the reconstructed image. As shown Fig. 3, DS-DMAS results in more noise reduction and reconstructed image has a higher quality in comparison with DAS and DMAS. To compare the sidelobe levels and noise reduction of proposed algorithm and two other beamformers, lateral variations in four depths of imaging are presented in Fig. 4. Lateral variation in depth of 25 $mm$, 30 $mm$, 35 $mm$ and 40 $mm$ are shown in Fig. 4(a), Fig. 4(b), Fig. 4(c) and Fig. 4(d), respectively. Clearly, sidelobe levels of DS-DMAS beamformer has the lowest level in comparison with DAS and DMAS, for instance, at the depth of 40 $mm$, sidelobe levels for DAS, DMAS and DS-DMAS are for about -20 $dB$, -40 $dB$ and -60 $dB$, respectively. Consequently, DS-DMAS results in 40 $dB$ and 20 $dB$ sidelobe levels reduction in comparison to DAS and DMAS, respectively.

To quantitatively compare the performance of the beamformers, signal-to-noise ratio ($SNR$) and full-width-half-maximum ($FWHM$) in -3 $dB$ are calculated in all depths of imaging using points positioned axially in all depths and 0 $mm$ lateral distance, shown in Fig. 1. The results are shown in Table I and Table II for $SNR$ and $FWHM$, respectively. $SNR$ is calculated using following equation:

$$SNR = 20\log_{10} P_{signal}/P_{noise}. \quad (11)$$

where $P_{signal}$ and $P_{noise}$ are difference of maximum and minimum intensity of each region, and standard deviation of the region, respectively. Considering Table I, in all depths of imaging DMAS and DS-DMAS beamformers outperform DAS. In particular, the depth of 50 $mm$ where $SNR$ of DAS, DMAS and DS-DMAS is for about 30.1526 $dB$, 38.5532 $dB$ and 43.7768 $dB$, respectively. In other words, DS-DMAS results in higher $SNR$ for about 13.6242 $dB$ and 5.2236 $dB$ in compare to DMAS and DAS, respectively. The outperforming of DS-DMAS beamformer in all depths of imaging is clear by considering presented results in the Table I. To evaluate the performance of proposed method in case of $FWHM$, consider Table II where results show that mentioned DMAS based algorithms outperform DAS, while DS-DMAS leads to narrower mainlobe width in all depths of imaging in compare to DMAS. For instance, consider depth of 55 $mm$ where DAS, DMAS and DS-DMAS leads to 2.2668 $mm$ 1.6922 $mm$ 1.2376 $mm$. It implies that resolution is improved using DS-DMAS in compare to DMAS and DAS. Moreover, the variation of $FWHM$, as long as depth of imaging increases, becomes less using DS-DMAS. In other words, differentiation of $FWHM$ in the first depth and the last depth of imaging results in 1.4055 $mm$, 1.1094 $mm$ and 0.8445 $mm$ for DAS, DMAS and DS-DMAS, respectively. Therefore, as long as depth of imaging increases, resolution degrades more slowly using DS-DMAS in compare to DMAS and DAS.

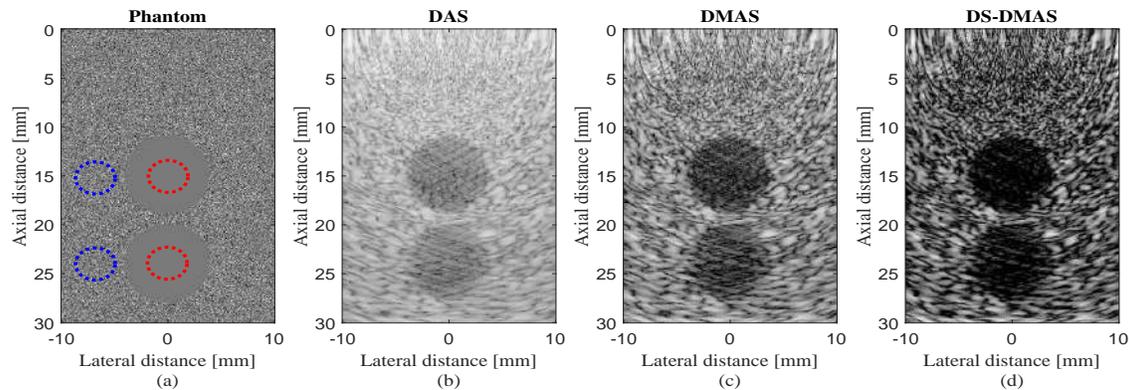

Fig. 5: Simulated cyst target using linear array. (a) Phantom, (b) DAS, (c) DMAS and (d) DS-DMAS. All images are shown with a dynamic range of 60 $dB$. 50 $dB$ noise was added to detected signals.

### B. Simulated Circular Cyst

Two cysts having radius of 4 $mm$ are located at depth of 15 $mm$ and 24 $mm$. Phantom of imaging, and reconstructed images using DAS, DMAS and DS-DMAS are shown in Fig. 5(a), Fig. 5(b), Fig. 5(c) and Fig. 5(d), respectively. The contrast ratio ($CR$) metric was used to quantitatively evaluate the performance of beamformers under cyst targets. $CR$ index is defined as:

$$CR = 20\log_{10}\left(\frac{\mu_{cyst}}{\mu_{bck}}\right) \quad (12)$$

where $\mu_{cyst}$ and $\mu_{bck}$ are the mean of image intensity before log compression inside the red and blue dotted circle in Fig. 5(a), respectively. Table III represents the calculated $CR$ using each beamformer for two cysts located at two depths. As can be seen in depth of 15 $mm$, DMAS and DS-DMAS beamformers lead to 12.5727 $dB$ and 21.591 $dB$ CR enhancement in compare to DAS, respectively. Moreover, DS-DMAS causes 9.0183 $dB$ CR enhancement in compare to DMAS. In the other hand, for cyst located at depth of 24 $mm$ DMAS-based algorithms outperform DAS beamformer while DS-DMAS results in 7.388 $dB$ improvement in $CR$ parameter in compare to DMAS beamformer.

TABLE III: **Contrast Ratio** ($dB$) Parameter For the Two Cysts Phantom.

| Beamformer | 15 $mm$ | 24 $mm$ |
|---|---|---|
| DAS | -14.9420 | -10.9521 |
| DMAS | -27.5147 | -20.0327 |
| DS-DMAS | -36.5330 | -27.4207 |

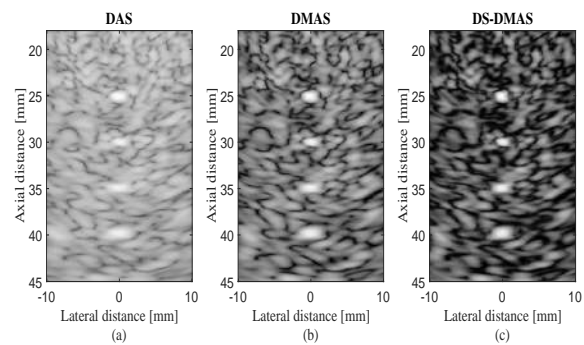

Fig. 6: Simulated point targets positioned in a low contrast background using linear array. (a) DAS, (b) DMAS and (c) DS-DMAS. All images are shown with a dynamic range of 60 $dB$. 50 $dB$ noise was added to detected signals.

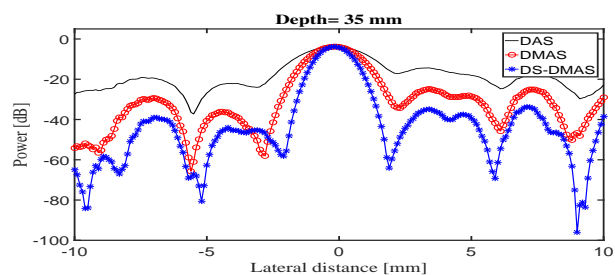

Fig. 7: Lateral variation of DAS, DMAS and DS-DMAS in depth of 35 $mm$.

### C. Low Contrast Target

In this section, DS-DMAS beamformer is evaluated using low contrast targets. Four 0.1 $mm$ spherical absorbers were positioned along the vertical axis every 5 $mm$ as initial pressure. The first absorber was 25 $mm$ away from the transducer surface. Imaging region was 20 $mm$ in lateral axis and 45 $mm$ in vertical axis. As can be seen in Fig. 6, point targets are more detectable in formed image using DS-DMAS while the background of imaging medium is retained. In order to compare beamformers in detail, lateral variation for each beamformer at the depth of 35 $mm$ is presented in Fig. 7. As is shown, DS-DMAS results in a narrower mainlobe. Moreover, DS-DMAS reduces the sidelobe levels for about 23 $dB$ and 11 $dB$ in comparison with DAS and DMAS, respectively.

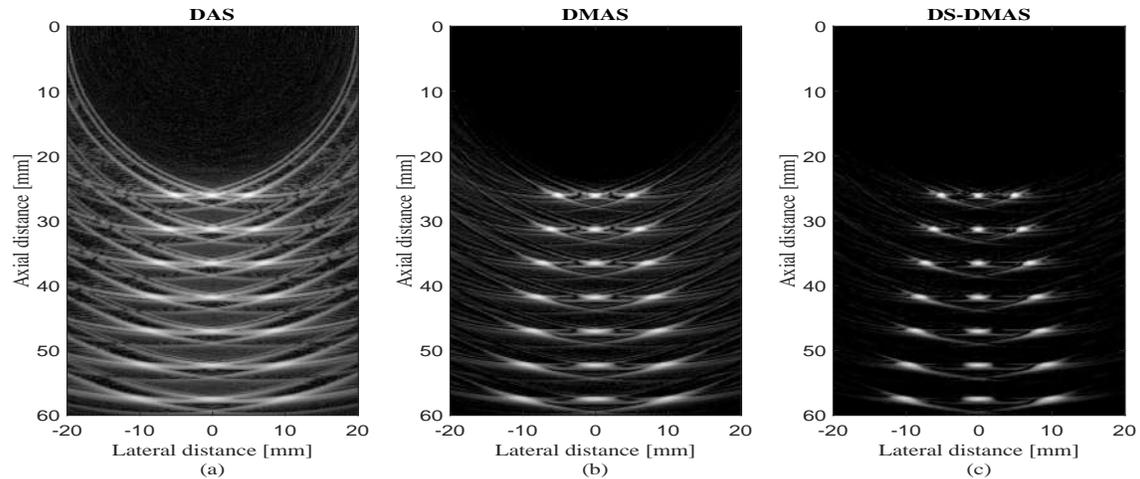

Fig. 8: Simulated three-point target using linear array. (a) DAS, (b) DMAS and (c) DS-DMAS. All images are shown with a dynamic range of 60 $dB$. 50 $dB$ noise was added to detected signals. The sound velocity is overestimated about 5%.

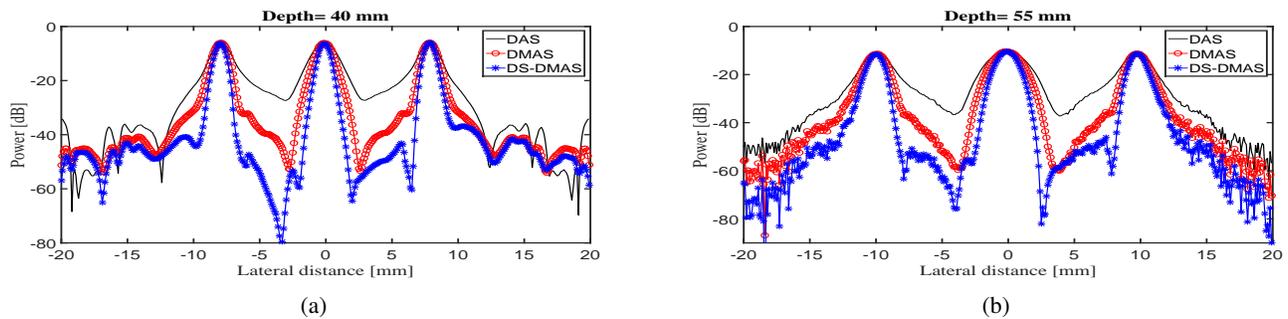

Fig. 9: Lateral variation of DAS, DMAS and DS-DMAS in depth of (a) 40 $mm$, (b) 55 $mm$. The sound velocity is overestimated about 5%.

### D. Sensitivity to Sound Velocity Inhomogeneities

In this section, the proposed method is evaluated in the term of robustness against the sound velocity errors resulting from medium inhomogeneities that are inevitable in practical imaging. The simulation design for Fig. 1 is used in order to investigate the robustness, except that the sound velocity is overestimated by 5%, which covers and may be more than the typical estimation error [38]. As can be seen in Fig. 8(a), the reconstructed image using DAS is suffering from negative effects of overestimated sound velocity. DMAS reduces the negative effects, and reconstructed image using DMAS, shown in Fig. 8(b), contains lower levels of artifacts. Even though the peak amplitudes are well estimated by DMAS, the negative effects are still annoying. As can be seen in Fig. 8(c), DS-DMAS reduces the effects of overestimated sound velocity in the reconstructed image and provides a higher image quality in comparison with DAS and DMAS. More importantly, the peak amplitudes are estimated the same as DAS and DMAS. Fig. 9 shows the lateral variation of the reconstructed images in Fig. 8. As can be seen, in the both presented depths, DS-DMAS results in the narrower mainlobe and lower levels of sidelobe in comparison with DAS and DMAS.

### E. Processing Complexity

In order to compare proposed method with other beamformers in term of computational burden, the number of operations needed for each algorithm are presented in Table IV. As can be seen, the order of computational complexity for DS-DMAS and DMAS is $O(M^2)$ which is exponentially more than $O(M)$ for DAS. This is the cost that is paid for higher performance of DMAS and DS-DMAS in comparison with DAS. On the other hand, the overhead computational cost of DS-DMAS in compare to DMAS, is linearly increased by increasing the number of employed elements of the transducer.

TABLE IV: **Processing Complexity for Different Beamformers**.

| Beamformer | Number of Operations |
|---|---|
| DAS | $M$ |
| DMAS | $\frac{M(M-1)}{2} + 2(M-1)$ |
| DS-DMAS | $M(M-1) + 3(M-1)$ |

## V. Experimental Results

To evaluate the DS-DMAS beamformer, in this section results of designed experiments are presented. A linear-array of PAI system was used to detect the PA waves and the major components of the system include an US data acquisition system, vantage 128 Verasonics (Verasonics, Inc., Redmond, WA), a Q-switched Nd:YAG laser with a pulse repetition rate of 25 $Hz$, wavelength 532 $nm$ and a pulse width of 10 $ns$. A transducer array (L7-4, Philips Healthcare) having 128 elements and 5.2080 $MHz$ central frequency was used as a receiver. A function generator is used to synchronize all operations (i.e., laser firings, PA signal recording). The sampling frequency on the receive is for about 20.8320 $MHz$ and the schematic of designed imaging system is presented in Fig. 10. Eight wires as absorbers were positioned along vertical axis, beginning 20 $mm$ from transducer surface, while each two wires at each depth are almost parallel. The system positioning is shown in Fig. 11 and it should be noticed that transducer surface is perpendicular to wires, so it is expected to provide a cross-section of the wires. The reconstructed images are shown in Fig. 12. As can be seen, DAS results in a low quality image with high level of artifacts, while DMAS enhances the reconstructed images. Although using DMAS results in an image with higher quality, it still suffers from artifacts. DS-DMAS leads to more enhanced image in compare to DAS and DMAS and also more noise reduction and levels of sidelobe degrading. To illustrate, consider the lateral variation of reconstructed images shown in Fig. 13, where lateral variations for two depths of imaging are presented. As can be seen in Fig. 13(a), sidelobe level using DAS is for about -30 $dB$, while using DMAS results in about 20 $dB$ sidelobe level reduction in compare to DAS and finally DS-DMAS leads to -65 $dB$ sidelobe level, which is the lowest sidelobe level among mentioned beamformers. Moreover, the valley of lateral variation in both presented depths of imaging has the lowest level using DS-DMAS. Consider, in particular, depth of 20 $mm$ shown in Fig. 13(a), where DAS, DMAS and DS-DMAS result in -40 $dB$, -52 $dB$ and -63 $dB$ valley of lateral variation, respectively. As a result, in both presented lateral variation, DS-DMAS results in the lowest level of sidelobes and artifacts, while point-targets are still detectable. Table V represents the calculated $SNR$ in two depths of imaging, shown by the yellow circles in Fig. 12. As can be perceived, DS-DMAS results in higher $SNR$ in comparison with DAS and DMAS. In particular, consider the depth of 20 $mm$ where DS-DMAS leads to $SNR$ improvement for about 23.1804 $dB$ and 11.9131 $dB$ in comparison to DAS and DMAS, respectively. In order to investigate the resolution in higher precision, another experiment has been developed. In Fig. 14, reconstructed images using two wires as targets are shown. As can be seen in Fig. 15(a) and Fig. 15(b), DS-DMAS results in a narrower mainlobe and lower levels of sidelobe in comparison with DAS and DMAS. $FWHM$ metric is presented in Table VI for two detected targets in Fig. 14. Consider, in particular, the depth of 30 $mm$ where DS-DMAS results in $FWHM$ improvement for about 0.9257 $mm$ and 0.4252 $mm$ in comparison with DAS and DMAS, respectively.

## VI. Discussion

The main enhancements gained by proposed beamforming algorithm are high noise reduction and artifact suppression. As can be seen in Fig. 1(a), Fig. 3(a), Fig. 8(a), Fig. 12(a) and Fig. 14(a), reconstructed images using DAS leads to high levels of sidelobe and artifacts which result from its inherent non-adaptiveness, blindness and high contribution of off-axis signals. Due to the combinatory coupling, multiplication and sum of the detected PA waves, reconstructed images using DMAS are improved in the cases of contrast resolution and contribution of noise. In other words, since DMAS algorithm calculates the auto-correlation of received PA signals, the output of this beamformer is the coherence of detected PA signals, and each calculated sample is weighted based on other calculated samples. This procedure of weighting makes DMAS away from blindness exists in DAS. It can be seen in Fig. 1(b), Fig. 3(b), Fig. 8(b), Fig. 12(b) and Fig. 14(b) that DMAS results in the higher quality images and more noise reduction. The contribution of off-axis signals leads to high levels of artifacts in the output of DAS beamformer while many of these suffering effects are addressed using DMAS and its correlation procedure. However, if the levels of artifacts increase, which mostly happens in high depths of imaging, or if imaging medium includes high levels of noise, the correlation procedure of DMAS will not be able to address the annoying effects. The effects of contribution of off-axis signals and high level of noise using DMAS can be seen in Fig. 1(b), Fig. 3(b) and Fig. 8(b), where quality of reconstructed point-target gets worse as depth of imaging increases. Also, it can be seen in Fig. 2 that although DMAS results in lower levels of sidelobe and a narrower width of mainlobe in compare to DAS, as depth of imaging increases the resolution and quality of reconstructed point-target get worse.

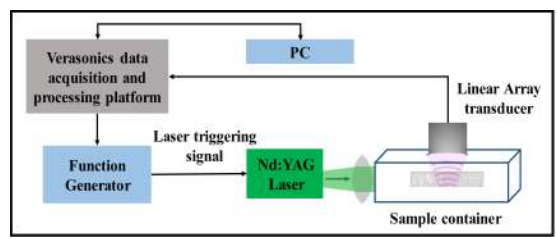

Fig. 10: Schematic of experimental setup.

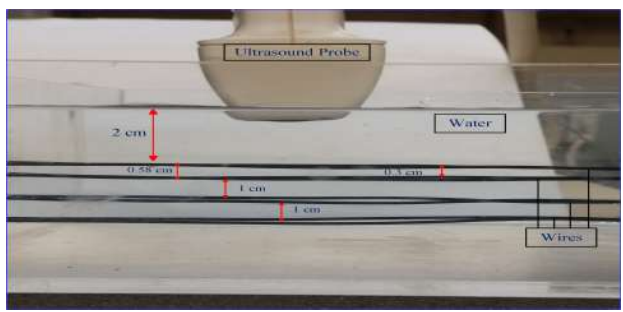

Fig. 11: Experimental setup of PA linear array imaging of eight wires.

TABLE V: **SNR** ($dB$) Values in Two Depths of Imaging.

| Beamformer | 20 $mm$ | 28 $mm$ |
|---|---|---|
| DAS | 32.9036 | 26.735 |
| DMAS | 44.1709 | 34.2061 |
| DS-DMAS | 56.0840 | 43.7240 |

TABLE VI: **FWHM** ($mm$) Values in Two Depths of Imaging.

| Beamformer | 30 $mm$ | 50 $mm$ |
|---|---|---|
| DAS | 1.4059 | 0.7384 |
| DMAS | 0.9054 | 0.4789 |
| DS-DMAS | 0.4802 | 0.3756 |

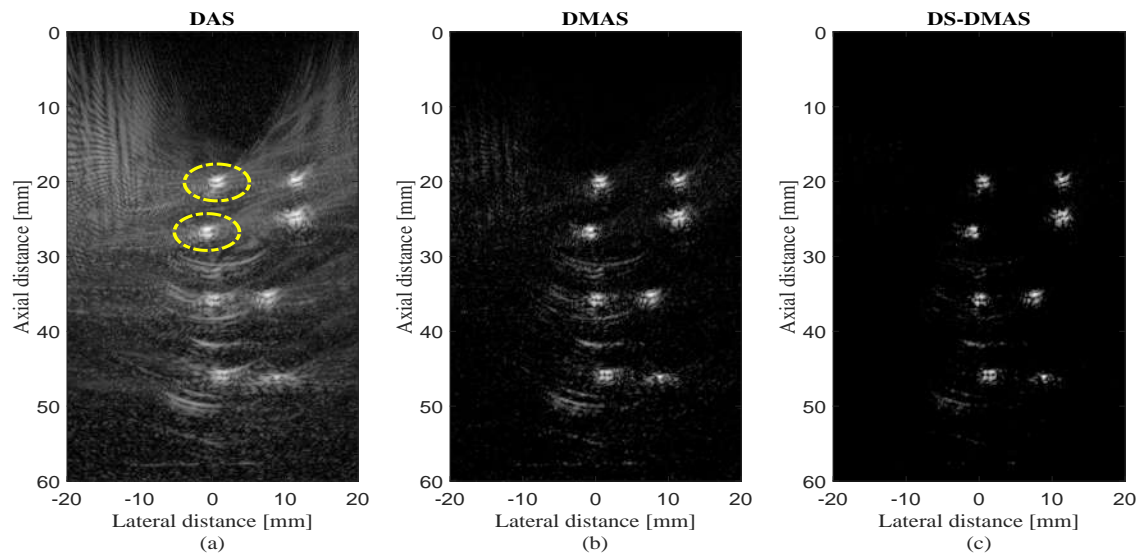

Fig. 12: Reconstructed experimental images using linear array. (a) DAS, (b) DMAS and (c) DS-DMAS. All images are shown with a dynamic range of 50 $dB$.

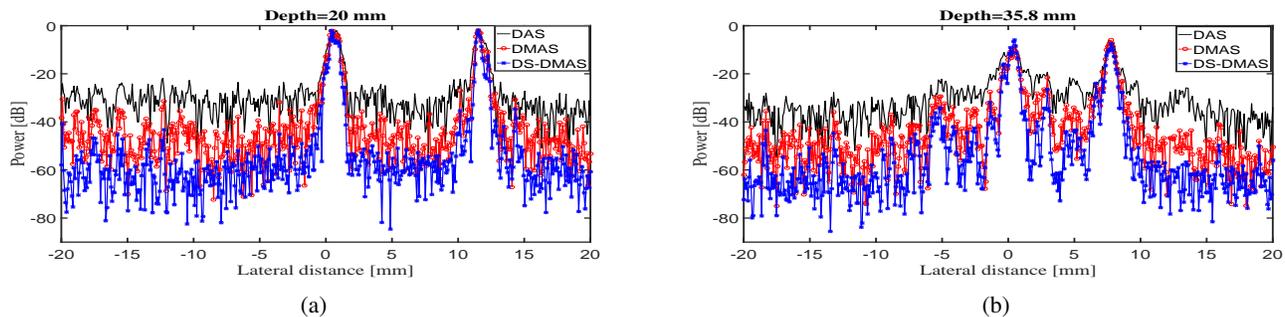

Fig. 13: Lateral variation of DAS, DMAS and DS-DMAS in depth of (a) 20 $mm$ and (b) 35.8 $mm$.

Expansion of DMAS beamformer algorithm is presented in (7) and (8), where there is a DAS between each term of the expansion. Consequently, if effect of off-axis signals and noise are presented, the quality of the reconstructed image will not be as expected. Since DMAS has been proved to be helpful in noise reduction and contrast resolution enhancement, this beamformer can be used instead of existing DAS inside DMAS algorithm expansion. To put it more simply, it is proposed to address the effects of high level of medium noise and contribution of off-axis signals using another correlation process inside DMAS. In DMAS, each calculated sample for each element is weighted based on detected signals using other elements of the array. The high level of medium noise and contribution of off-axis signals can prevent DMAS in (7) to weight samples correctly. DS-DMAS modifies the calculated weights using two stages of DMAS. As can be seen in Fig. 1(c) and Fig. 2, sidelobe levels and negative effect of DMAS are degraded using DS-DMAS and reconstructed point target are more detectable, especially in high depths of imaging. Considering Fig. 3(c) and Fig. 4, it can be perceived that DS-DMAS results in more enhanced image in the presence of high level of medium noise in comparison to DMAS and DAS. All simulations and calculated parameters indicate that DS-DMAS beamformer outperforms DMAS and DAS in all parameters at the expense of more number of operations in compare to DMAS and DAS. Of note, it is expected that DS-DMAS uses the strong signals and discards the weak signals.

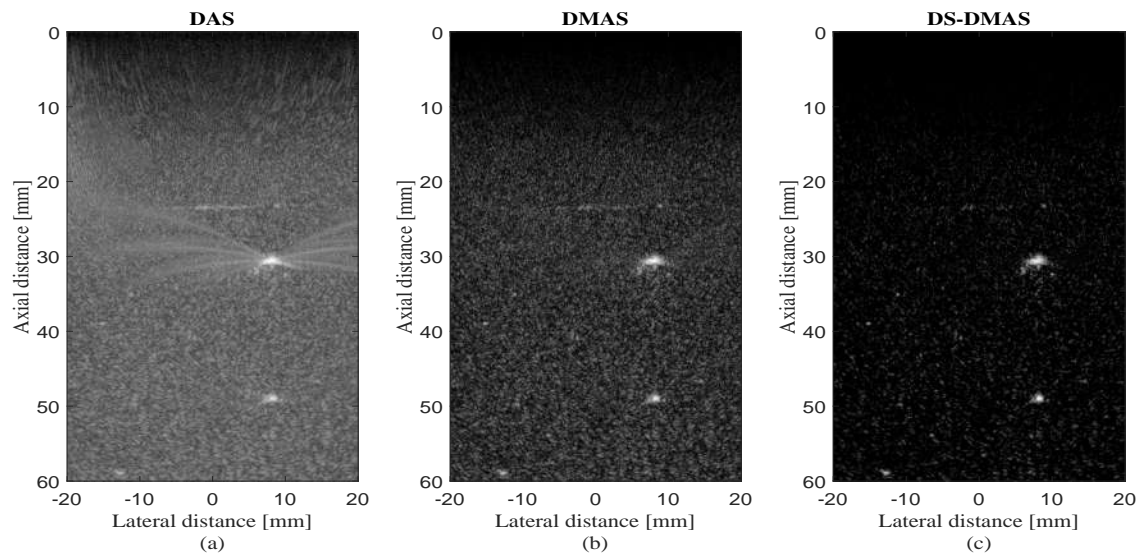

Fig. 14: Reconstructed experimental images using linear array. (a) DAS, (b) DMAS and (c) DS-DMAS. All images are shown with a dynamic range of 60 $dB$.

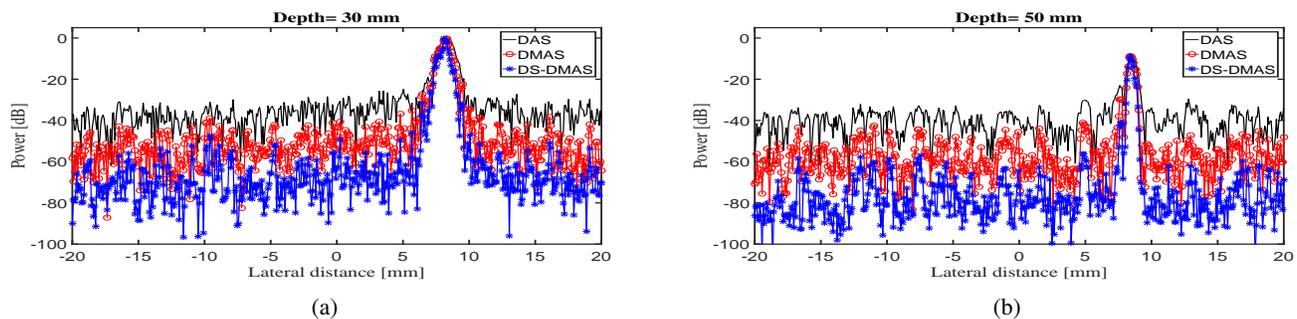

Fig. 15: Lateral variation of DAS, DMAS and DS-DMAS in depth of (a) 30 $mm$ and (b) 50 $mm$.

In Fig. 6, performance of DS-DMAS has been evaluated using point targets having low contrast in comparison with the imaging medium. Based on visual evaluation and presented lateral variation in Fig. 7, DS-DMAS provides images with higher quality in comparison to DAS and DMAS in case of low contrast targets. In order to investigate the robustness of proposed method, sound velocity has been overestimated in reconstruction procedure. Formed images are shown in Fig. 8, and corresponding lateral variations in two depths of imaging are presented in Fig. 9. DS-DMAS estimates the peaks amplitude correctly while reduces the negative effects of overestimated sound velocity. Reconstructed images using experimental data are shown in Fig. 12 and Fig. 14. As can be seen, reconstructed images using DS-DMAS contains lower levels of sidelobe and artifacts. In other words, DS-DMAS retains strong signals while degrading the weak signals in order to form images with higher quality in compare to DMAS and DAS. Lateral variations of experimental reconstructed images are presented in Fig. 13 and Fig. 15, which show DS-DMAS outperforms DAS and DMAS in the terms of levels of sidelobe. Moreover, the $SNR$ metric for experimental images, shown in Fig. 13, are presented in Table V and show the superiority of DS-DMAS in comparison with DAS and DMAS. Apart from that, resolution improvement can be seen in Fig. 15, along with corresponding $FWHM$ metric presented in Table VI.

## VII. Conclusion

In PAI, DAS beamformer the most common image reconstruction algorithm due to its simple implementation, formed image suffers from poor resolution and high levels of sidelobe. To overcome these limitations, DMAS algorithm was proposed, which was recently used in US imaging. DMAS beamformer is based on the correlation of detected signals at the elements of the array and expanding DMAS formula leads to multiple terms of DAS. In this paper, we introduced a modified version of DMAS beamforming algorithm, called DS-DMAS. This algorithm is based on existing DAS in the expansion of DMAS algorithm and it is proposed to use DMAS instead of existing DAS in the expansion. It has been shown that using DMAS-based algorithms leads to sidelobes levels reduction and resolution improvement in comparison with DAS, while DS-DMAS outperforms DMAS. Numerical simulations using three-point target at different depths of imaging and two levels of noise, along with cyst target and experimental results are presented. Results are discussed for each beamforming algorithm in terms of sidelobe levels, resolution. The qualitative and quantitative comparisons show that DS-DMAS beamformer significantly reduces levels of sidelobe for about 15 $dB$ and improves resolution. Moreover, DS-DMAS leads to improvement in terms of $SNR$, $FWHM$ and $CR$ for about 13%, 30% and 35% in compare to DMAS, respectively.